# Temporal graphs

Vassilis Kostakos
Department of Mathematics and Engineering
University of Madeira

and

Human-Computer Interaction Institute
Carnegie Mellon University

`vassilis@cmu.edu`

**Abstract**: We introduce the idea of temporal graphs, a representation that encodes temporal data into graphs while fully retaining the temporal information of the original data. This representation lets us explore the dynamic temporal properties of data by using existing graph algorithms (such as shortest-path), with no need for data-driven simulations. We also present a number of metrics that can be used to study and explore temporal graphs. Finally, we use temporal graphs to analyse real-world data and present the results of our analysis.

# Introduction

Graphs and complex networks have been used to study many complex human and natural phenomena [e.g. 8,9,16,20]. Typically, graph structures are used to represent relationships between entities such as individuals or organisations. While these relationships are mostly instantiated intermittently over time, previous research has used network representations that aggregate the relationships at discrete intervals [e.g. 17,19]. Hence we refer to such graphs as "static" graphs.

Aggregating data to derive snapshots at distinct intervals makes data analysis tractable, but this approach is also historically motivated by the fact that rich temporal information was not traditionally available for analysis. To overcome this lack of rich temporal data, Granovetter and Schelling were among the first to propose a *simulation* approach to studying dynamic processes such as diffusion on static graph structures [7,18]. Given a static structure, they effectively proposed a linear threshold model where, at each simulated time-step, each node becomes "active" if a certain fraction of its neighbours is already active. This approach helps us overcome possible lack of rich temporal data, and has become widely popular in subsequent research [3,11,17,21,22].

Not until recently have digital and communication technologies enabled us to capture on an unprecedented scale data about many aspects of human behaviour, such that its temporal richness is adequately preserved. For instance, researchers have captured and analysed extensive longitudinal data on people's use of the phone [15], people's mobility around a city [6,14] or a campus site [1,13], and longitudinal interactions between group members [4,5]. The availability of such data has enabled researchers to employ *emulations*, as opposed to simulations, for studying diffusion and propagation in complex and social networks [10,12,23].

## Emulations

Emulation-based analysis is a data driven approach for understanding dynamic behaviour. For instance, consider a dataset consisting of the emails that a group of people exchange over time. Using this dataset we can represent each person as an agent, and then run through the following process: for each event (in this case an email exchange) decide whether one agent "infects" the other. By adjusting the various diffusion parameters and observing the effect we gain insights on how this group of people may share information over time, or how viruses may spread through a community.

One of the main advantages of using emulations is that its results are arguably more realistic than an arbitrary simulation. Obviously this realism depends on the quality and quantity of data. Additionally, emulations are a good tool for exploring the effect of real-world alternative scenarios (e.g. what will happen if the email server goes offline for 2 days?).

However, the use of emulations has its drawbacks. Most notably, emulations sacrifice the key benefit of static graph analysis: deriving concrete yet universal metrics for each node or unit of interest. This means that we can no longer apply our understanding of well-studied graph metrics. Furthermore, emulation analyses are more "messy" and harder to extrapolate from, precisely because they are data driven. It is much more difficult to clearly and accurately externalise an emulation dataset as opposed to a graph structure.

These difficulties make it hard to compare across emulation datasets and application domains. To overcome these difficulties, a possible compromise is to use the approach of graph snapshots at distinct intervals [e.g. 17,19]. This, however, will sacrifice most of the rich temporal information in the dataset.

## Contribution

In this paper we describe *temporal* graphs, a tool for analysing rich temporal datasets that describe events over periods of time. Temporal graphs have the analytical benefits of static graph analysis while at the same time retain all temporal information that may be available to us. Additionally, we define a number of metrics for temporal graphs, namely temporal proximity, geodesic proximity, and temporal availability, all of which help us quantify the relationship between nodes over time, and the role of each node in the temporal context of the entire network.

In the next section we introduce the idea of temporal graphs by demonstrating their construction with a small sample dataset. We explore the same dataset using two different sets of assumptions: first we assume that our data was generated using technology that allows for one-way communication (e.g. email), and then we follow the same process for two-way communications (e.g. telephone).

Finally, we use temporal graphs to analyse real world datasets of social interactions. We find that our temporal metrics are distinct from static graph metrics and uniquely quantify people's relationship over time. Additionally, we show how our metrics allow us to quickly identify key nodes in dynamic processes.

## Temporal graphs

Let us assume a dataset describing events that take place at distinct points in time. We also assume that each distinct event does not have a temporal duration. For instance, consider a dataset consisting of the email exchanges between a group of users. Each email exchange takes place at a specific moment, and additionally each exchange is instantaneous. Furthermore, this is a one-way communication channel, with information flowing unidirectionaly from the sender to the recipient. For example, in Table 1 we show such a sample dataset consisting of email exchanges between 5 people over a period of 21 days. For each email we know the sender, recipient(s) and the day on which the email was sent.

| Sender | Recipient | Time |
|--------|-----------|------|
| A | B | $t_1=0$ |
| A | C,E | $t_2=1$ |
| E | D | $t_3=3$ |
| B | C | $t_4=5$ |
| B | D | $t_5=9$ |
| D | B | $t_6=14$ |
| A | D | $t_7=20$ |

Table 1 : The email dataset: list of emails exchanged between people at distinct points in time (days).

At this stage, our use of days as a unit of measurement is arbitrary, as hours or seconds can equally be used for the same purpose. What is important to note is that we assume each email transaction has no duration in itself.

## Static representation

If we ignore all temporal information from Table 1, we can construct a graph to represent the aggregated relationships between the five people. We proceed by creating one node per person, and linking nodes that had an email exchange. The direction of links is the same as the direction of the respective email message. The resulting social network is shown in Figure 1.

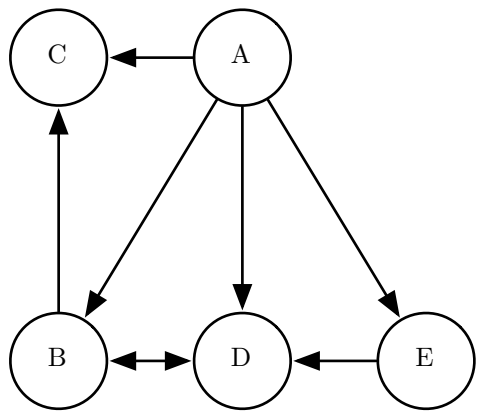

Figure 1 : A static graph representation of Table 1, which retains no temporal information. Each node represents a person, while edges link people who had email exchanges (in the direction of communication).

The graph in Figure 1 enables us to calculate a number of distinct metrics for each node as well as for the graph as a whole. For instance, we may consider each node's centrality (degree, closeness, betweenness) as a mechanism for interpreting our data. However, we point out that this representation has already sacrificed all temporal information that was available in Table 1, such as the frequency of events and the time difference between subsequent events. As far as Figure 1 is concerned, all links are simultaneously and continuously available.

## Temporal graphs and metrics

To retain the temporal information of Table 1 in a graph we construct a temporal graph in three steps:

(1) Create one node per person per point in time. Hence person A is represented by the set of instances $\{At_1, At_2, At_7\}$.

(2) For each set of instances we link consecutive pairs $\{At_x, At_{x+1}\}$ with directed edges of weight $t_{x+1} - t_x$, representing the temporal distance between the pair. For example, the weight between nodes $At_2$ and $At_7$ is 19 days (20-1).

(3) We use unweighted directed edges to link node instances that participated in a email transaction. An email from A to B at time $t_x$ is instantiated as a directed link between $At_x$ and $Bt_x$.

Hence, each node in Figure 1 now becomes a directed chain of nodes that represent all temporal instances of the node over time. Following these conventions, we produce the temporal graph in Figure 2.

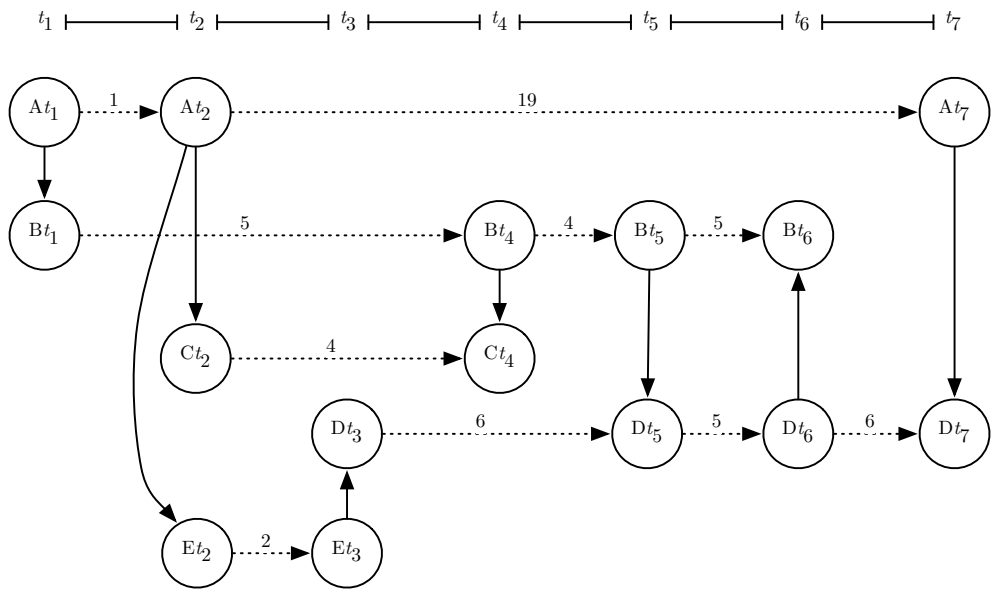

Figure 2: A temporal graph representation of Table 1. For readability, "waiting" links are dashed. Solid links represent instantaneous email transactions, and carry no weight.

The temporal graph in Figure 2 has much richer information about the events described in Table 1, as well as the flow of information that these events enable. For readability, all weighted links are dashed, while solid links have no associated weight.

We can already observe some insights that temporal graphs offer. For example, in Figure 1 there is a path from D to C via B (DBC). This is quite misleading however, as in Figure 2 we see that no such path exists. To be precise, there is no path from *any* instance of D to *any* instance of C. This discrepancy between Figures 1 and 2 arises because the interaction between D and B (e.g. $Dt_6 - Bt_6$) takes place after B interacts with C ($Bt_4 - Ct_4$), hence D cannot reach C at all. Similarly we observe that many paths in Figure 1 are not available in Figure 2, such as paths DBC, EDBC, and ADB.

*Temporal proximity*

We observe that in Figure 1, person A has three paths to D: AD, ABD, AED. While AD is the shortest in Figure 1, this is not necessarily the case in Figure 2. To demonstrate this, first we need to define a metric of distance for temporal graphs. Since Figure 2 has no single node for person A or D, but rather has the sets $\{At_1, At_2, At_3, At_7\}$ and $\{Dt_3, Dt_5, Dt_6, Dt_7\}$ respectively, we need to define a measure of distance between these two sets.

We call our metric of distance *temporal proximity* between X and Y, defined as

$$p(X,Y,t_a,t_b)$$

where $t_a$ is a temporal pre-condition for X and $t_b$ is a temporal post-condition for Y. The values $t_a$ and $t_b$ can either take a specific value or be empty (*null*). This gives rise to at least four possible ways of calculating temporal proximity, all of which can be calculated using any weighted shortest-path algorithm. For instance, to measure the temporal proximity between A and D, we may choose to calculate:

- $p(A,D,t_i,t_j)$: the shortest path between some instance of A, e.g. $At_2$, and some instance of D, e.g. $Dt_7$. This example is equivalent to $p(A,D,t_2,t_7)$ and intuitively means "Given $t_2$, find the shortest path from A to D such that D is reached at $t_7$". In this case, the shortest path has weight $w(At_2,At_7,Dt_7) = 17$, which translates back to AD.

- $p(A,D,t_i,null)$: the shortest path between some instance of A, e.g. $At_1$, and any instance of D ($\{Dt_3, Dt_5, Dt_6, Dt_7\}$). This example is equivalent to $p(A,D,t_1, null)$, and intuitively means "Given time $t_1$, find the shortest path from A to D". In this case, the shortest path is AED with total weight $w(At_1,At_2,Et_2,Et_3,Dt_3) = 1+0+2+0=3$.

- $p(A,D,null,t_i)$: the shortest path between any instance of A ($\{At_1, At_2, At_3, At_7\}$) and some instance of D, e.g. $Dt_5$. This example is equivalent to $p(A,D,null,t_5)$ and intuitively means "Find the shortest path from A to D such that D is reached at $t_5$". In this case, there are two shortest paths with weight $w(At_1\ At_2\ Et_2\ Et_3\ Dt_3\ Dt_5) = w(At_1\ Bt_1\ Bt_4\ Bt_5\ Dt_5) = 9$, which translate back to AED and ABD respectively.

- $p(A,D,null,null)$: the shortest path between any instance of A ($\{At_1, At_2, At_3, At_7\}$) and any instance of D ($\{Dt_3, Dt_5, Dt_6, Dt_7\}$). This intuitively means "find the shortest possible path from A to D throughout the entire dataset". In this case, the shortest path is $w(At_7,Dt_7) = 0$, which translates back to AD.

Our notion of temporal proximity $p$ inevitably must take into account time. While for a static graph we can simply calculate the geodesic distance between two nodes in terms of hops, in a temporal graph we need to effectively set the time limits within which this path is to be instantiated, hence the need for temporal pre- and post-conditions in defining $p$. As a result, the shortest paths in Figures 1 and 2 are quite different. Most notably, as we pointed out in the previous section, the existence of a path in the static graph (Figure 1) does not guarantee the existence of a path in the temporal graph: we see that while a path exists from A to D in Figure 1, $p(A,D,t_3,t_6)$ has no solution, and hence evaluates to *null*.

*Average temporal proximity*

A further concept we introduce is *average temporal proximity*, defined as

$$P(X,Y) = \Sigma p(X,Y,t_i,null)/n, \, p(X,Y,t_i,null) \neq null$$

where $P(X,Y)$ measures "on average, the time it takes to go from X to Y". If any $p(X,Y,t_i,null) = null$, then we ignore it and decrease $n$ accordingly. This means that $P$ is not affected by temporarily unavailable paths. For example,

$$P(A,D) = (p(A,D,t_1,null) + p(A,D,t_2,null) + p(A,D,t_7,null)) / 3$$

$$= (3 + 2 + 0) / 3 = 1.67$$

We also define $P_{in}$ and $P_{out}$ as

$$P_{in}(X) = \Sigma P(i,X)/n, i \neq X, P(i,X) \neq null$$

$$P_{out}(X) = \Sigma P(X,i)/n, i \neq X, P(X,i) \neq null$$

where $P_{in}(X)$ and $P_{out}(X)$ are a measure of "on average, how quickly is X reached by the rest of the network" and "on average, how quickly does X reach the rest of the network" respectively. These are simply the column and row averages of $P$ discarding the diagonal and *null* values.

In Table 2 we calculate $P$, $P_{in}$ and $P_{out}$ for our data. The unit of measurement for $P$ is time, e.g. days.

|   | \multicolumn{5}{c}{P} | $P_{out}$ |
|---|---|---|---|---|---|---|
|   | A | B | C | D | E |   |
| A | 0 | 6.5 | 0.5 | 1.67 | 0.5 | 2.29 |
| B | - | 0 | 2.5 | 4.33 | - | 3.42 |
| C | - | - | 0 | - | - | - |
| D | - | 5.33 | - | 0 | - | 5.33 |
| E | - | 12 | - | 1 | 0 | 6.5 |
| $P_{in}$ | - | 7.94 | 1.5 | 2.33 | 0.5 |   |

Table 2 : P, Pin and Pout for people using email. Unit is days.

*Geodesic proximity*

Our next metric is *geodesic proximity*, defined as

$$g(X,Y,t_a,t_b)$$

denotes the least number of hops between X and Y given temporal pre-condition $t_a$ for X and temporal post-condition $t_b$ for Y. This measure discards the weights on the temporal graph, yet it is still subject to the temporal restrictions imposed by the unidirectional waiting links (dashed links). Borrowing our earlier examples on temporal proximity, we see that:

- $g(A,D,t_1,null) = 3$ (i.e. $At_1, At_2, At_7, Dt_3$),
- $g(A,D,null,t_5) = 5$ (i.e. $At_1\ At_2\ Et_2\ Et_3\ Dt_3\ Dt_5$),
- $g(A,D,t_2,t_7) = 2$ (i.e. $At_2, At_7, Dt_7$),
- $g(A,D,null,null) = 1$ (i.e. $At_7, Dt_7$).

The temporal pre- and post-conditions $t_a$ and $t_b$ operate in exactly the same way as in the case of calculating temporal proximity $p$.

*Average geodesic proximity*

In addition to calculating the geodesic proximity between nodes, we can also calculate the *average geodesic proximity* between nodes, defined as

$$G(X,Y) = \Sigma g(X,Y,t_i,null)/n, g(X,Y,t_i,null) \neq null$$

where $G(X,Y)$ is a measure of "on average, how many hops is X away from Y", hence discarding weighs on edges but retaining edge directionality. As a concrete example,

$$G(A,D) = (g(A,D,t_1,null) + g(A,D,t_2,null) + g(A,D,t_7,null)) / 3 =$$
$$(3 + 2 + 1) / 3 = 2.$$

We also define $G_{in}$ and $G_{out}$ as

$$G_{in}(X) = \Sigma G(i,X)/n, i \neq X, G(i,X) \neq null$$

$$G_{out}(X) = \Sigma G(X,i) /n, i \neq X, G(X,i) \neq null$$

where $G_{in}(X)$ and $G_{out}(X)$ are a measure of "on average, in how many hops is X reached by the rest of the network" and "on average, in how many hops does X reach the rest of the network" respectively. It may be argued that we should not count hops between instances of the same person, as that represents "waiting" and does not really involve transmission of anything. However, we have decided to retain such "waiting" hops in our measurements, because they represent distinct events in time and opportunities that arise. We return to this point during our discussion.

In Table 3 we show $G$, $G_{in}$ and $G_{out}$ for our data.

|   | G |   |   |   |   | $G_{out}$ |
|---|---|---|---|---|---|---|
|   | A | B | C | D | E |   |
| A | 0 | 3.5 | 1.5 | 2 | 1.5 | 2.125 |
| B | - | 0 | 1.5 | 2 | - | 1.75 |
| C | - | - | 0 | - | - | - |
| D | - | 2 | - | 0 | - | 2 |
| E | - | 4.5 | - | 1.5 | 0 | 3 |
| $G_{in}$ | - | 3.33 | 1.5 | 1.83 | 1.5 |   |

Table 3 : G, Gin and Gout for people using email. Unit is hops.

In Table 3 the unit of measurement is hops (or temporal events). We see that each node is always 0 hops away from itself. We note that AB is 3.5 hops, despite the fact that AB are a single hop away in a static graph representation (Figure 1).

*Temporal availability*

To calculate $P$ and $G$ we discard any path with $p = null$ and $g = null$ respectively. Since $P$ and $G$ are average values, we lose information about how many of the relative paths where actually available. For instance, $P(B,C) = 2.5$ which is relatively small even though of the four instances of B only two can actually reach any instance of C. To get further insights about the temporal relationships between nodes we introduce the concept of *temporal availability*, defined as

$$V(X,Y) = size \{g(X,Y,t_i,null) = null\} / n$$

where $V$ is a measure of the probability that there exists *any* path between X and Y at any given moment. Note that instead of $g$ we can use $p$ with identical results. We also define $V_{in}$ and $V_{out}$ as

$$V_{in}(X) = \Sigma V(i,X)/n, i \neq X, V(i,X) \neq null$$

$$V_{out}(X) = \Sigma V(X,i) /n, i \neq X, V(X,i) \neq null$$

where $V_{in}(X)$ and $V_{out}(X)$ measure "on average, what is the probability that the network can reach X" and "on average, what is the probability that X can reach the network". From Figure 2 we see that

$$V(A,D) = size\{g(A,D,t_1,null), g(A,D,t_2,null), g(A,D,t_7,null)\} / 3 = 1,$$

and

$$V(B,C) = \text{size } \{g(B,C,t_1,null), g(B,C,t_4,null)\} / 4 = 0.5.$$

In Table 4 we show $V$, $V_{in}$ and $V_{out}$ for our data. We observe that each node is (obviously) always available to itself (probability 1). We also observe that the pairs with $V=0$ are also pairs that had null $G$ and $P$ in Tables 2 and 3.

|   | \multicolumn{5}{c}{V} | $V_{out}$ |
|---|---|---|---|---|---|---|
|   | A | B | C | D | E |   |
| A | 1 | 0.67 | 0.67 | 1 | 0.67 | 0.75 |
| B | 0 | 1 | 0.5 | 0.75 | 0 | 0.31 |
| C | 0 | 0 | 1 | 0 | 0 | 0 |
| D | 0 | 0.75 | 0 | 1 | 0 | 0.19 |
| E | 0 | 1 | 0 | 1 | 1 | 0.5 |
| $V_{in}$ | 0 | 0.6 | 0.29 | 0.68 | 0.17 |   |

Table 4 : V, Vin and Vout between people using email. Unit is probability.

## Temporal graphs and bidirectional data

We now revisit our dataset from Table 1 and slightly reinterpret it. Let us assume that instead of email, the communications in Table 1 took place using an two-way technology such as the telephone. This reinterpretation of the data is shown in Table 5.

| Caller | Callee | Time |
|---|---|---|
| A | B | $t_1=0$ |
| A | C,E | $t_2=1$ |
| E | D | $t_3=3$ |
| B | C | $t_4=5$ |
| B | D | $t_5=9$ |
| D | B | $t_6=14$ |
| A | D | $t_7=20$ |

Table 5 : The phone dataset: list of telephone calls between people at distinct points in time. In case of multiple callees, we assume the caller has a direct line to each of the callees, but the callees are not directly connected.

Once again we construct a static graph representing information in an aggregated form. In Figure 3 we create one node per person, and link nodes that spoke together on the phone. The result is similar to Figure 1, except that all edges are bidirectional.

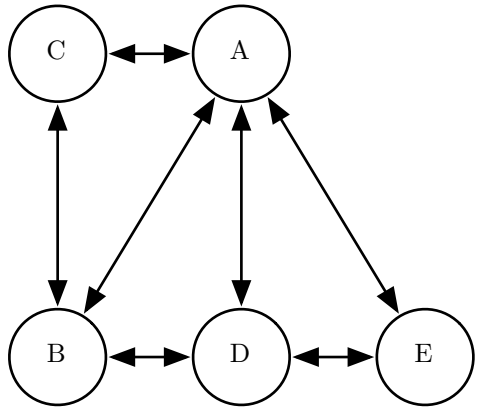

Figure 3 : A graph representation of Table 5, produced by discarding all temporal information. Here, each node represents a person, while links denote people who spoke on the phone.

Next, we generate a temporal graph following the procedure described earlier. The only difference is that in step (3) we create bidirectional links to represent phone calls, as opposed to unidirectional links representing emails. The result is shown in Figure 4.

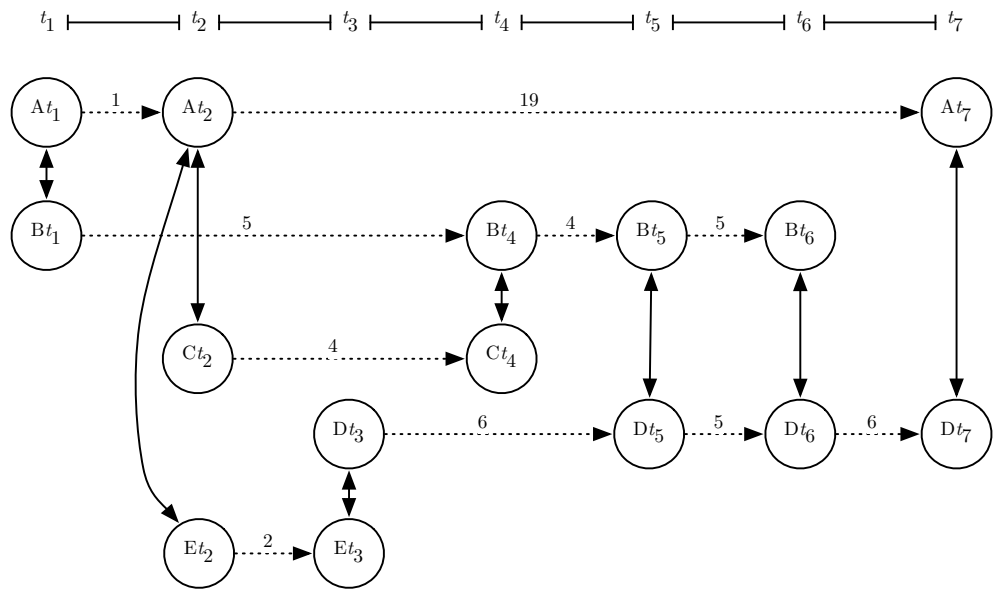

Figure 4: A temporal graph representation of Table 5. For clarity, each "waiting" link is dashed. Solid edges represent telephone conversations, and carry no weight.

In case of multiple callees, we have not created a direct link between the callees themselves in order to preserve our semantics. Hence, $Ct_2$ and $Et_2$ are not linked directly although their temporal proximity $p(C,E,t_2,t_2)$ is 0.[1]

---

[1] In this case, we assume that A initiated calls to C and E on two separate lines, so that C and E can only communicate with each other via A, albeit instantaneously.

In Tables 6 - 8 we show *P*, *G*, and *V* for our data

|   | \|  | *P* |   |   |   | $P_{out}$ |
|---|---|---|---|---|---|---|
|   | A | B | C | D | E |   |
| A | 0 | 2 | 0.5 | 1.67 | 0.5 | 1.67 |
| B | 8 | 0 | 0.5 | 1.75 | 1 | 2.82 |
| C | 7.5 | 2 | 0 | 3 | 0 | 3.13 |
| D | 8.5 | 2 | - | 0 | 0 | 3.5 |
| E | 8.5 | 5 | 0 | 1 | 0 | 3.63 |
| $P_{in}$ | 8.13 | 2.75 | 0.33 | 1.85 | 0.36 |   |

Table 6 : P, Pin and Pout between people using the telephone. Unit is days.

|   | \|  | *G* |   |   |   | $G_{out}$ |
|---|---|---|---|---|---|---|
|   | A | B | C | D | E |   |
| A | 0 | 2 | 1.5 | 2 | 1.5 | 1.75 |
| B | 3.25 | 0 | 1.5 | 1.75 | 3 | 2.36 |
| C | 3.5 | 1.5 | 0 | 3 | 2 | 2.5 |
| D | 2.5 | 1.33 | - | 0 | 1 | 1.61 |
| E | 3 | 3.5 | 2 | 1.5 | 0 | 2.5 |
| $G_{in}$ | 3.06 | 2.08 | 1.67 | 1.06 | 1.86 |   |

Table 7 : G, Gin and Gout between people using the telephone. Unit is hops.

|   | \|  | *V* |   |   |   | $V_{out}$ |
|---|---|---|---|---|---|---|
|   | A | B | C | D | E |   |
| A | 1 | 0.67 | 0.67 | 1 | 0.67 | 0.75 |
| B | 1 | 1 | 0.5 | 1 | 0.25 | 0.69 |
| C | 1 | 1 | 1 | 1 | 0.5 | 0.86 |
| D | 1 | 0.75 | 0 | 1 | 0.25 | 0.5 |
| E | 1 | 1 | 0.5 | 1 | 1 | 0.86 |
| $V_{in}$ | 1 | 0.85 | 0.42 | 1 | 0.41 |   |

Table 8 : V, Vin and Vout between people using the telephone. Unit is probability.

## Analysis of real-world temporal datasets

We now use temporal graphs and their metrics to study real-world datasets. Here we use the analytic tools we developed in the previous section to explore the dynamic behaviour and properties of the datasets. We consider the Enron email corpus,[2] and data on people's face to face encounters from the Cityware project [14]. We analyse each dataset using both static graphs and temporal graphs.

---

[2] Retreived from http://www-2.cs.cmu.edu/~enron/ on July 10, 2008.

## Data

The datasets we consider are:

(1) The collection of emails sent within the Enron corporation during October 1999. For each email we know the sender and recipient(s), and the exact date and time the message was sent. In constructing a temporal graph for this dataset we used unidirectional links to denote emails.

(2) The set of face to face encounters between people, as recorded by a Cityware scanner at a pedestrian walkway in the University of Bath during the first half of March 2008. This data is collected by means of Bluetooth technology, which records a unique ID for each person, and the date and time when two people were in very close range (up to 10m) of each other. In our temporal representation we use bidirectional links to denote face to face encounters between people.

Table 9 has more details about each dataset, while a visual representation of our data is shown in Figure 5.

|  | Dataset | |
|---|---|---|
|  | Enron | Cityware |
| Data points | 3747 | 23705 |
| Duration | 30 days | 15 days |
| Static graph |  |  |
|   Nodes | 944 | 881 |
|   Edges | 3747 | 23705 |
| Temporal graph |  |  |
|   Nodes | 3603 | 24697 |
|   Edges | 6406 | 71226 |

Table 9 : Details for our datasets, and their respective static and temporal graphs.

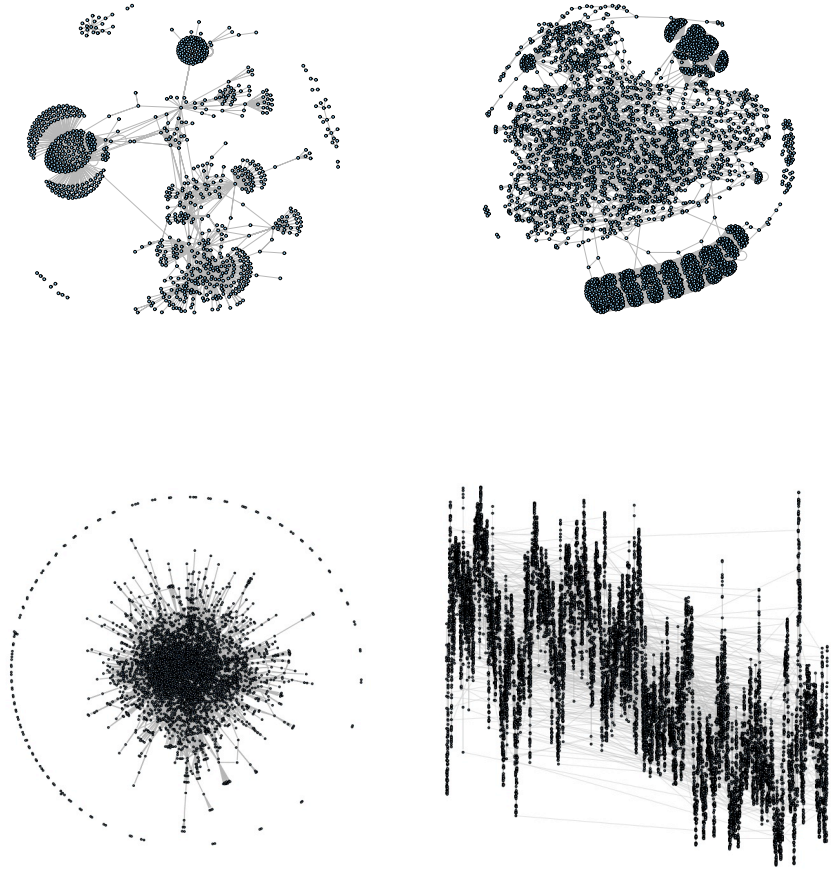

Figure 5: The left column shows the static graph representations of our data, the right column shows the temporal graphs. (top: the enron dataset, bottom: the Cityware dataset). All layouts derived using the same algorithm (Fruchterman-Reingold).

## Results

In Table 10 we show *P*, *G* and *V* for each dataset. Here we have averaged the values for ever pair of nodes in each dataset, ignoring *null* values.

In Figure 6 we show the degree distribution for each static graph (left). Additionally, we show the distribution of instance set size (Figure 6 middle), which is a measure of how many nodes in the temporal graph represent each node from the static graph. This data is taken from step (1) of constructing a temporal graph. Also, we show the distribution of link weights for each temporal graph (Figure 6 right).

In Figure 7 we show the relationships between $G_{in}$ - $G_{out}$, $P_{in}$ - $P_{out}$, $V_{in}$ - $V_{out}$ for each dataset (top: Enron, bottom: Cityware). Additionally, we have colour-coded each data point according to the corresponding node's degree in the static graphs. Hence, red point are nodes with high degree in the static graph, while nodes with lowest degree appear in blue, with yellow, green and orange colours in between. Following the same colour coding, in Figure 8 we show the relationship between $V_{in}$ - $P_{in}$, $V_{out}$ - $P_{out}$, $V_{in}$ - $G_{in}$, and $V_{out}$ - $G_{out}$.

Finally, in Figure 9 we show histograms of P for both datasets. Note that the x-axes in this figure are in days, while the y-axes are frequency.

|  | P | G | V |
|---|---|---|---|
| Enron | 9.86 (6.45) days | 23.65 (15.19) hops | 0.007 (0.07) |
| Cityware | 3.05 (2.59) days | 15.52 (0.43) hops | 0.64 (0.43) |

Table 10 : Average P, G and V values for our datasets, and standard deviation in brackets.

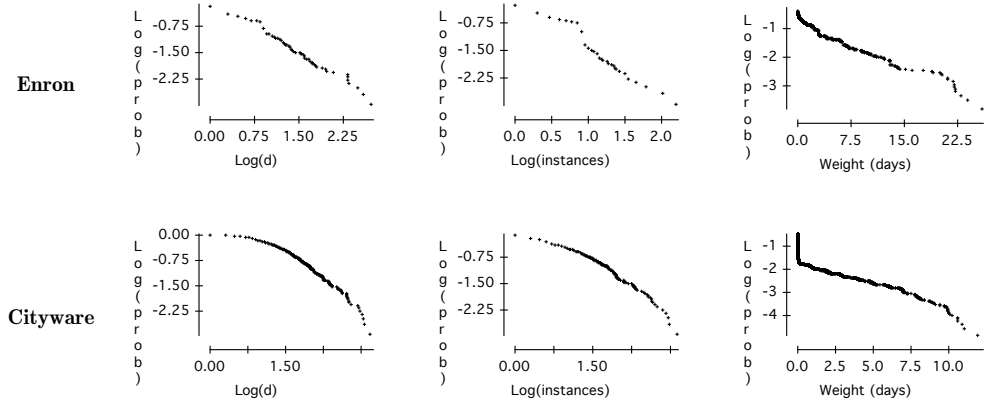

Figure 6 : For each dataset we show the distribution of degree (log-log plot), instance set size (log-normal) and link weight (log-normal). Degree distribution is calculated for static graphs. All y-axes show cumulative probability $p(X \geq x)$.

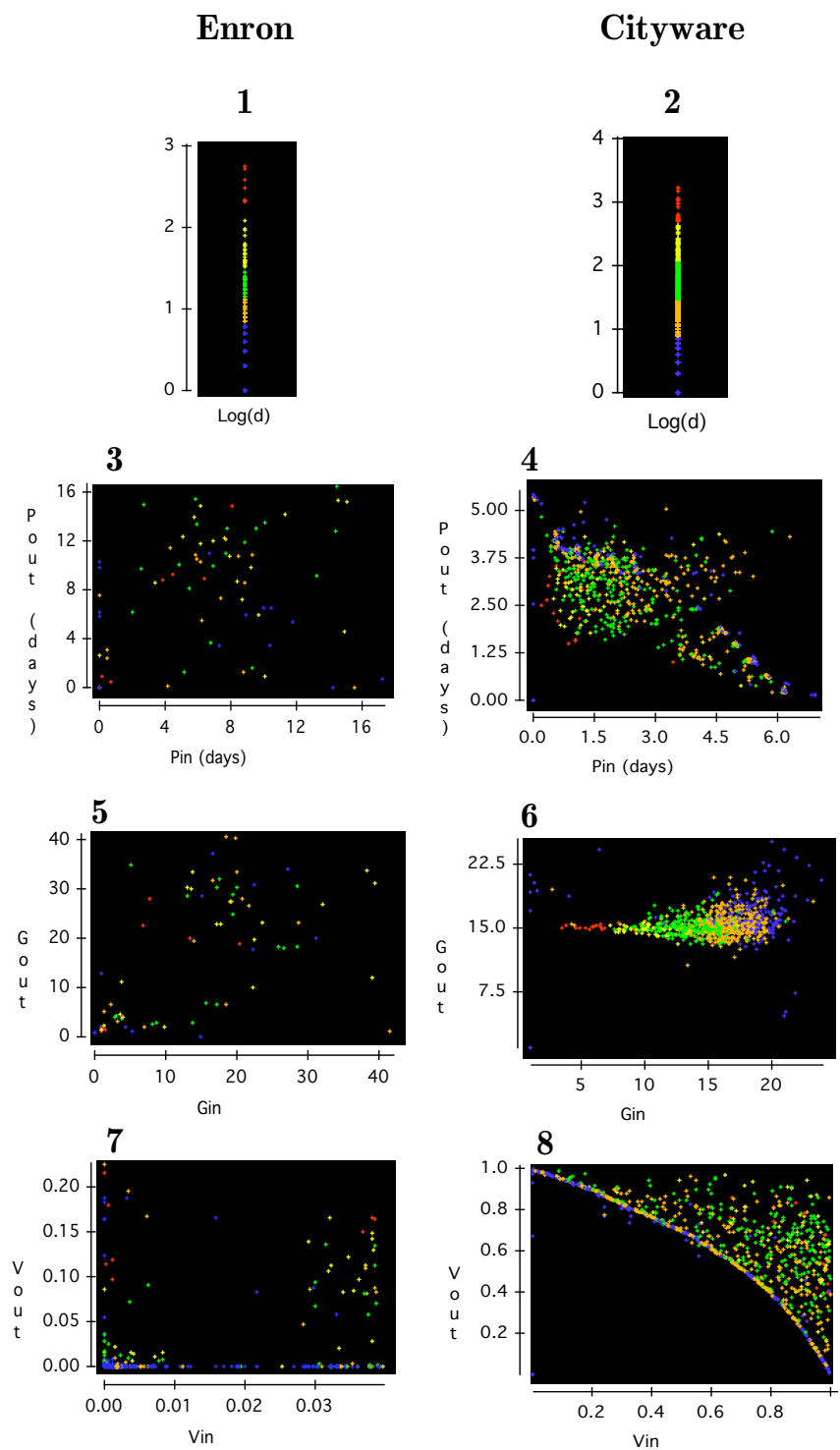

Figure 7 : Correlation between Gin - Gout, Pin - Pout, and Vin - Vout. Left column is the Enron dataset, right column is the Cityware dataset. The data is colour-coded according to degree d (calculated on the static graphs).

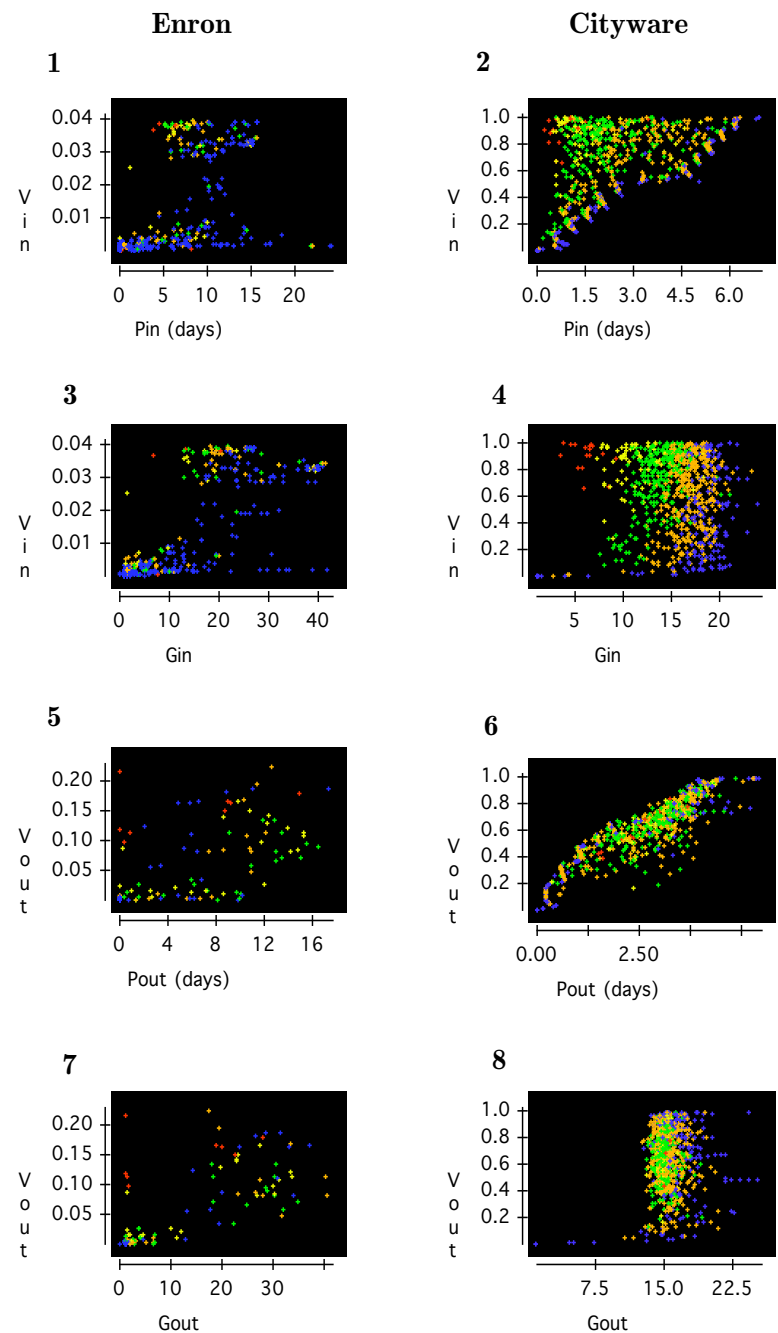

Figure 8 : Correlation between Vin - Pin, Vin - Gin, Vout - Pout, and Vout - Gout. Left column is the Enron dataset, right column is the Cityware dataset. The data is colour-coded according to Figures 7.1 & 7.2.

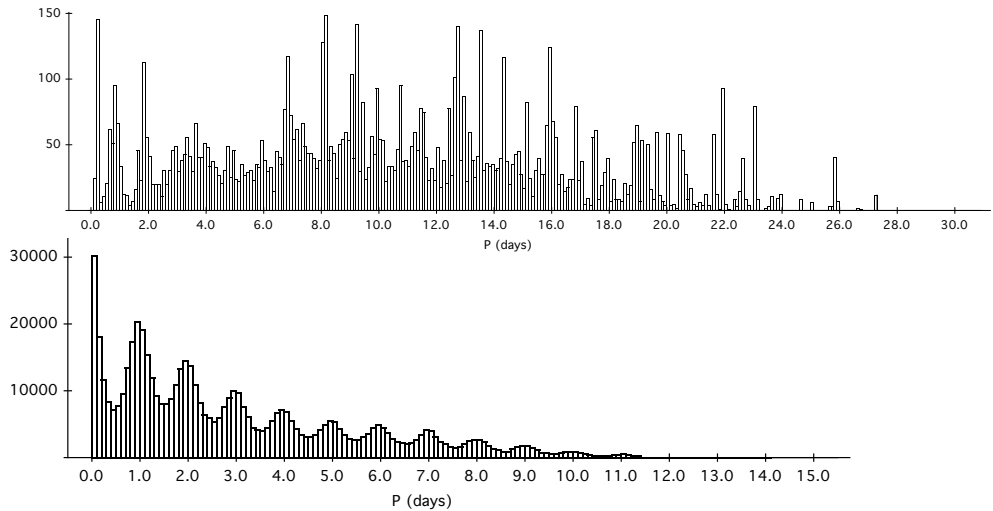

Figure 9 : Histogram of P for both datasets (top: Enron, bottom: Cityware). Y-axis is frequency, x-axis is days.

## Discussion

In this paper we introduced temporal graphs, and demonstrated how to construct them using a minimal dataset with either unidirectional or bidirectional data (e.g. email vs. phone). In each case we show how to derive various node metrics, including the average temporal proximity *P,* average geodesic proximity *G,* and temporal availability *V*. Furthermore, we analysed two real-world datasets using our metrics, and we now discuss the types of insights that our analysis has enabled.

Temporal graphs are a tool for understanding the dynamic properties of a dataset and the corresponding entities and relationships it represents. In considering the temporal dynamics of a dataset, *P* is a measure of how quickly one person can reach another in terms of time, *G* give us insights into the number of hops/events/opportunities that take place before one person reaches another, while *V* is the probability that a node can reach some other node at any given time. Depending on the domain, one measure may be of greater interest that another. For example, when considering epidemics and diffusion we are more likely to be interested in the amount of time it takes for someone to be infected (*P*). On the other hand, if we are considering the spread of information through an opportunistic mobile ad-hoc network of sensors and devices, *G* may be of more importance because it represents the number of decisions that the forwarding algorithm needs to make (e.g. decide to keep the information on the current device, or decide to transmit the information to a nearby device), which has repercussions for the performance of the system as a whole.

Similarly, $P_{in}$ and $P_{out}$ measure a person's relationship with the rest of the network in terms of absolute time, $G_{in}$ and $G_{out}$ give us insight into the person's relationship with the network in terms of hops, while $V_{in}$ and $V_{out}$ are the actual probability that a person can reach or can be reached by the network as a whole. Hence, all these metrics help us assess the suitability of nodes for receiving or broadcasting information over time.

In our analysis of the Enron and Cityware data we wish to better understand how information flows, or can flow, in these networks over time. In the case of Enron we assume that the information flows via emails, while for the Cityware data we assume that information can flow either through people's face to face encounters or through their respective mobile devices. Understanding how people's mobile devices can opportunistically spread and receive information is part of an ongoing research towards building mobile peer-to-peer applications for data exchange.

From Table 10 we see that the Cityware community is much more "dense" and "tighter" than the Enron community. The people in the Cityware dataset are both quicker in reaching each other and require far less hops than the Enron individuals, even though the high standard deviation for Enron's values suggests that some individuals are very well connected. What is striking however is the extremely low $V$ for Enron, suggesting that only a small amount of temporal paths are actually available on average. Hence we can argue that information does not flow as well between the Enron individuals.

### Structure

In terms of structure, Figure 6 interestingly suggests that the two datasets are quite similar. The degree distribution d in both static graphs follows an approximate power law as we see in Figure 6 (left). This is to a large extent an expected result as similar scale-free features have been observed before when analysing human relationships [e.g. 20].

A further similarity between the two datasets is in the distribution of instance set size. Recall that when constructing a temporal graph, each node becomes a set of instances representing the node over the course of time. In Figure 6 (middle) we show just how many instances did we have to create in this process. For both datasets the set size follows an approximate exponential decay (log-normal plot). This quantity represent the number of times a particular person appears in the dataset, hence it is a good indication of how many opportunities that person will have to receive or transmit information.

Finally, in the right of Figure 6 we show the link weight distribution which again follows an approximate exponential decay. This quantity represents the amount of time between subsequent appearances of a specific node, hence the frequency with which a node is active in the network. These metrics suggest similarities in the way both networks are structured: the number of links that each person establishes in each datasets, the number of times each person appears in the dataset and the time between subsequent appearances all follow a very similar distribution.

### Within-metrics analysis

To explore further the reason why given our datasets' structural similarities the average values for $P$, $G$ and $V$ vary so much, we turn to Figure 7. Here we split each of $P$, $G$ and $V$ in terms of their *in* and *out* components (top: Enron, bottom: Cityware). Each data point represents a person, and we have colour coded all points according to the number of connections (i.e. an indication of centrality), ranging from red to yellow, orange and blue (Figures 7.1 and 7.2). In terms of $P$ we see that in the Enron dataset some people are very quick in reaching the network and being reached by the network (bottom left corner 7.3). It is interesting to note, however, that some red data points appear in the top of 7.3, suggesting individuals highly connected locally but slow in reaching the whole network. Similarly in this graph we can identify locally low-connected individuals who can nevertheless reach the network very quickly.

Considering $P$ for Cityware (7.4) we identify roughly two clusters of people. In addition to the main cluster we see a cluster extending towards the bottom right corner of the graph, itself made of a series of smaller clusters. These are individuals who can very quickly reach the network, but are slow to be reached. Furthermore we observe that data points' colour roughly changes as we move from the bottom-left towards the top-right, hence suggesting that as nodes gain more links they are quicker to reach the network and quicker in being reached.

A comparative analysis between 7.3 and 7.4 suggests that the Enron dataset is much less structured, having people who are extremely quick in reaching the network and being reached by the network, but also highly-connected individuals who are extremely slow at propagating information through the network. On the other hand, the cityware data suggests that the variation between people is much smaller. This distinction may also reflect the difference between a business environment where things need to get done quickly by small number of key people, as opposed to the Cityware data which reflects the relatively quiet and time-tabled University environment. In the latter case a small number of people are highly connected because they regularly visit or pass through the physical area where the data was collected.

Considering the average geodesic proximity of Figures 7.5 and 7.6 we once again observe the messiness of the Enron dataset as opposed to the structure of Cityware. In 7.5 we observe that many individuals can reach and be reached by the network in very few steps, even less than the best-performing individuals in the Cityware data. It is interesting to note that in 7.6 the average path of nodes reaching to the network is about 15 hops with relatively little variability. This is not observed when we look at the length of incoming links ($G_{out}$), where in addition the colour of data points is a very good indicator of how easily a person can be reached by the network.

Finally, the temporal availability $V$ highlights a striking breakdown in the Enron dataset: half the people are very hard to reach, while the other half are relatively easier to reach (7.7). And once again we observe a split between the highly-connected (red) nodes: half are hard to reach, half are not. On the other hand, 7.8 suggests that in the Cityware dataset there is an apparent inverse relationship between nodes' ability to reach the network and being reached by the network, mostly followed by low-degree nodes (blue and orange). These represent people who were seen relatively few times in our dataset. If such a person was observed near the start of our observation, they would appear in the top-left corner of the graph: they can reach many subsequent devices, but they cannot be reached since they never reappear. If on the other hand such a person was seen towards the end of our observation, then they would be good at receiving information from the network but bad at reaching any of the previous seen people, hence they would tend towards the bottom-right hand corner of 7.8. As a person starts to appear more often (e.g. green, yellow or red) then they are increasing their chances of being able to send and receive data from then network.

## Between-metrics analysis

We now consider our datasets in terms of between-metrics analysis. We explore our two datasets by looking at how temporal availability $V$ relates to how quickly and in how many hops individuals can communicate. We already saw in Figure 7.7 that the Enron dataset is quite abruptly split between individuals who can be reached with relatively high probability and individuals who are most likely unreachable. We explore this dichotomy further in Figures 8.1 and 8.3 where we consider its relationship to temporal and geodesic proximity $P_{in}$ and $G_{in}$. In these figures we observe an "hourglass" silhouette, suggesting that despite their differences in incoming availability, the amount of time it takes to reach individuals varies both within the high availability and low availability groups. The same holds for geodesic proximity, where we find low-availability individuals being both very close and very far from the rest of the network (bottom-left and bottom-right in 8.3).

A similar analysis for the Cityware data yields completely different results. Here we observe that highly-degree individuals (red) are both highly-available and quickly reached (top-left in 8.2) as well as reached in a small number of hops (top-left in 8.4). For low-degree individuals (blue & orange) we observe a linear relationship between the probability that they can be reached and the amount of time in which they can be reached. As such, those who are on average quickly reached have relatively few incoming temporal paths available. On the other hand, those with slightly greater $P_{in}$ are much more likely to be reachable. Once again, this can be attributed to low-degree individuals who appear early or late in the dataset. High-degree individuals on the other hand tend to cluster in the top-left corner of 8.2, suggesting both high availability and relatively short temporal proximity.

It is interesting to note the massive shift in the data point going from 8.2 to 8.4. Most data points have shifted to the right, such that colour becomes a very good predictor of how easily nodes can be reached from the rest of the network (we also observed this in 7.6). Hence we find that for low-degree nodes a large number of hops is required before they can be reached -- even though this happens in a relatively short period of time. From 8.2 and 8.4 we conclude that high-degree individuals are very good information receptors as they can be reached quickly and easily from the rest of the network, and with very high availability. As nodes connectivity is gradually reduced however, the time it takes to reach them and the number of hops very quickly increase. This is not the case in the Enron dataset, where even low-degree individuals are still relatively quickly and easily reached.

Next, we consider individuals' ability to transmit information to others in the network. In the Enron dataset we found that many individuals had a much higher chance of reaching others, i.e. high $V_{out}$ in 8.5 and 8.7, as opposed to being reached by others. Furthermore we note that the amount of time and the number of hops individuals need to reach others is relatively consistent. In the Cityware dataset, we observe a strong linear relationship between the probability of being able to reach someone and the number of days this will take (8.6). In the same figure we note that most high-degree individuals score relatively low in being able to find a path to transmit information, albeit when they find one they are able to reach the others quite quickly. However the whole dataset is shifted to the right when we consider the geodesic proximity for transmitting information to the network. In this case we observe that most paths are at least 15 hops long (something we also verify in 7.6). This seems to be a lower bound which limits all individuals, even those with high probability of finding a path to their recipient.

**Frequency analysis**

In Figure 9 we have plotted histograms of $P$ for both datasets. Here we examine every single pair of nodes in each dataset, and calculate its temporal proximity. We then bin our all values as seen in the two histograms in Figure 9. It is important to note the following crucial detail about our histograms. The x-axis is real time, hence the interval between the 0 and 1 marks represents 0 to 24 hours, the interval between the 1 and 2 marks represents 24 to 48 hours, and so on.

A similarity across both histograms is that there are local maxima near full integer values, and local minima near 0.5 values. In other words we observe relatively more pairs of individuals having temporal proximity that is the multiple of full day, while relatively fewer have temporal proximity that includes half-days. We argue that this is evidence of the expression "daily routine" taken literally. Here we observe that communication and activity takes place in daily "waves": people reading email and responding, people visiting the university and going to work or class. From the histograms we see that if a temporal path does exist between two nodes, then it is most likely to "ride the wave" of daily routine and be instantiated when most activity takes place.

Despite these similarities, however, the two histograms exhibit a striking difference in terms of regularity. The Enron dataset (top) is highly erratic as opposed to the wave-like distribution of the Cityware data. We see that in the Cityware dataset if a temporal path exists between two people it is much more likely to be quick. Furthermore, the probability of a temporal path being long gradually decreases at its length increases. On the other hand, we observe distinct peaks throughout the Enron dataset, with many temporal paths being up two 3 weeks long.

### Implications

There are a number of implications arising from the metrics and insights we have derived from our temporal graph analysis. First, we are able to use our metrics to compare relative data sets and understand the similarities and differences in people's temporal behaviour. We found that the Enron dataset is much more erratic, messy, and extreme, while the Cityware data was underpinned by consistency and routine. This was despite the fact that both datasets exhibit very similar structural properties.

In addition, our analysis can also be used to understand the different role of individuals within their network, their potential for sending and receiving information, and the network's ability to propagate this information in general. For instance, if we are interested in diffusion processes within these networks, we know that individuals who are well-positioned to receive information quickly are not necessarily good at finding the temporal paths to propagate this information through the network. Specifically, in the Enron dataset it is even possible that well-connected individuals may actually be worse at disseminating information than others.

Furthermore our graphs are very good as a tool for comparing diffusion speeds through networks. For instance, we see in Figure 7.4 that within 3 days the bulk of individuals in the Cityware dataset will have access to the diffused information, while in the case of the Enron dataset the respective duration is about 12 days. Yet we observed that Cityware dataset had a hard bound of 15 hops required before one can be reached, while within the Enron dataset there are potentially much shorter paths.

This kind of analysis is also useful in developing and optimising communication systems that take advantage of the opportunistic behaviour of human contact and interaction. Such ad-hoc peer to peer networks work best when we can reliably identify individuals who have a good chance of forwarding information to the ultimate recipient both quickly and in the least number of hops. Our analysis in Figure 8 demonstrates the relationship between speed (in terms of both time and hops) versus chance of success. These may be used as a basis in developing forwarding algorithms, or expressing the inherent differences between two different sample populations.

## Conclusion and ongoing work

In this paper we have presented temporal graphs which are a graph representation that can retain rich temporal information about the underlying temporal dynamics. A key strength of graphs in general is their amiability for use in communicating and describing to others our data, as well as for deriving concrete universal metrics that are well studied and understood across domains.

Temporal graphs offer a basis for obtaining the same benefits when dealing with inherently dynamic data, both for describing and communicating the data itself, as well as for analysing and understanding its properties. In this paper we present an array of metrics that can be used to characterise a temporal dataset, and use these to compare two distinct real-world datasets. A key benefit of temporal graphs is that rely on standard shortest-path algorithms, hence most existing software tools can easily cope with temporal graph analysis.

A key assumption we have made in this paper is that temporal events have no duration in themselves. While this may be an appropriate assumption for email communication, the same is not necessarily true for face-to-face communications and other domains where the concept of temporally overlapping events is crucial. We are currently working towards developing the necessary tools for being able to express event duration in our temporal graphs.

# Acknowledgments

The author thanks Alan Penn, Eamonn O'Neill and Jose Luis. This work is supported by the Portuguese Foundation for Science and Technology via the CMU-Portugal agreement.

# References

1. Balazinska M, Castro P. (2003). Characterizing mobility and network usage in a corporate wireless local-area network. In: MobiSys '03 proceedings first international conference on mobile systems, applications and services. ACM Press, New York, pp 303–316
2. Barabasi, A.-L. and Albert R. (1999). Emergence of scaling in random networks Science, 286 509–512.
3. Berger, E. (2001). Dynamic Monopolies of Constant Size. Journal of Combinatorial Theory Series B 83, 191-200.
4. Chaintreau A., Hui P., Crowcroft J., Diot C., Gass R., Scott J. (2006). Impact of human mobility on the design of opportunistic forwarding algorithms. In: Proceedings 25th IEEE conference on computer communications (INFOCOM). IEEE CS Press, New York
5. Eagle N., Pentland A. (2006). Reality mining: sensing complex social systems. Pers Ubiquitous Comput 10(4):255–268
6. Gonzalez,M.C., C.A. (2008). Hidalgo and A.-L. Barabási. Understanding individual human mobility patterns Nature 453, 479-482.
7. Granovetter, M. (1978). Threshold models of collective behavior. American Journal of Sociology 83(6): 1420-1443.
8. Holme, P. (2003). Network dynamics of ongoing social relationships. Europhysics Letters, 64, 3, 427-433.
9. Holme, P., Park, S.M., Kim, B.J., Edling, C.R. (2007). Korean university life in a network perspective: Dynamics of a large affiliation network. Physica A, 307, 821-830.
10. Kostakos, O'Neill and Penn. (2007) Brief encounter networks. arXiv 0709.0223.
11. Macy, M. Chains of Cooperation: Threshold Effects in Collective Action. American Sociological Review 56(1991). M. Macy, R. Willer. From Factors to Actors: Computational Sociology and Agent-Based Modeling. Ann. Rev. Soc.
12. McNamara, L., Cecilia Mascolo and Licia Capra. (2008). Media Sharing based on Colocation Prediction in Urban Transport. To appera in Proceedings of ACM International Conference on Mobile Computing and Networking (Mobicom08). San Francisco, CA, USA.
13. McNett M., Voelker G.M. (2005). Access and Mobility of Wireless PDA Users. SIGMOBILE Mob Comput Commun Rev 9(2):40–55
14. O'Neill, E., Kostakos, V., Kindberg, T., Fatah gen. Schiek, A., Penn, A., Stanton Fraser, D. and Jones, T. (2006). Instrumenting the city: developing methods for observing and understanding the digital cityscape. In proceedings of UbiComp 2006, Lecture notes in Computer Science 4206, Springer, pp. 315-332.
15. Onnela, J.-P., J. Saramäki, J. Hyvönen, G. Szabó, D. Lazer, K. Kaski, J. Kertesz, and A.-L. Barabási (2007). Structure and tie strengths in mobile communication networks, PNAS 104, 7332-7336.
16. Onody, R.N. and de Castro, P.A. (2004). Complex network study of Brazilian soccer players. Physical Review E, 037103.
17. Sarkar,P., and Moore, A. (2005). Dynamic Social Network Analysis using Latent Space Models. SIGKDD Explorations: Special Editionon Link Mining.
18. Schelling, T. (1978). Micromotives and Macrobehavior. Norton.
19. Snijders, T.A.B. (2001). The statistical evaluation of social network dynamics. In Sociological Methodology (M.E. Sobel and M.P. Becker Eds), Boston and London: Basil Blackwell, 361-395.
20. Strogatz, S.H. (2001). Exploring complex networks. Nature, 410, 268-276.
21. Valente, T. (1995). Network Models of the Diffusion of Innovations. Hampton Press.
22. Watts, D. A. (2002). Simple Model of Global Cascades in Random Networks. Proc. Natl. Acad. Sci. 99, 5766-71.
23. Yoneki, E., and Crowcroft, J (2008). Wireless Epidemic Spread in Dynamic Human Networks. To appear in Bio-Inspired Computing and Communication, LNCS 5151, Springer.

# Appendix - Generating temporal graphs

## Code for generating temporal graphs

```php
<?php
$file = @fopen('data.txt','r');
$last_seen["dummy"] = 11;   // instantiate global variable
$two_way = false;           // are we using 1- or 2-way technology? Fig2 vs. Fig4

function attach_instance($device,$date){
    // See if a device has appeared previously.  If so, create a directed link
    // from previous instance to this instance.  The weight of the link is the
    // time difference between the two instances.

    global $last_seen;
    $previous_date = $last_seen[$device];

    if(($previous_date != "") && ($previous_date != $date)){
       $diff = $date - $previous_date;
       echo $device.$previous_date."  \t  ".$device.$date."  \t  ".$diff."\n";
    }
    $last_seen[$device] = $date;
}

while(!feof($file)){
   $line      = fgets($file);
   $items     = explode(",", $line);
   $date      = trim($items[0]);
   $sender    = trim($items[1]);
   $recepient = trim($items[2]);

   attach_instance($sender,$date);
   attach_instance($recepient,$date);
   echo $sender.$date."  \t  ".$recepient.$date."  \t  0\n";
   if($two_way == true){
      echo $recepient.$date."  \t  ".$sender.$date."  \t  0\n";
   }
}
?>
```

## Sample input & output

| Input file (see Tables 1 & 5) | Ouput for Figure 2: | | | Output for Figure 4: | | |
|---|---|---|---|---|---|---|
| 0,A,B  | A0  | B0  | 0  |     |     |    |
| 1,A,C  | A0  | A1  | 1  | A0  | B0  | 0  |
| 1,A,E  | A1  | C1  | 0  | B0  | A0  | 0  |
| 3,E,D  | A1  | E1  | 0  | A0  | A1  | 1  |
| 5,B,C  | E1  | E3  | 2  | A1  | C1  | 0  |
| 9,B,D  | E3  | D3  | 0  | C1  | A1  | 0  |
| 14,D,B | B0  | B5  | 5  | A1  | E1  | 0  |
| 20,A,D | C1  | C5  | 4  | E1  | A1  | 0  |
|        | B5  | C5  | 0  | E1  | E3  | 2  |
|        | B5  | B9  | 4  | E3  | D3  | 0  |
|        | D3  | D9  | 6  | D3  | E3  | 0  |
|        | B9  | D9  | 0  | B0  | B5  | 5  |
|        | D9  | D14 | 5  | C1  | C5  | 4  |
|        | B9  | B14 | 5  | B5  | C5  | 0  |
|        | D14 | B14 | 0  | C5  | B5  | 0  |
|        | A1  | A20 | 19 | B5  | B9  | 4  |
|        | D14 | D20 | 6  | D3  | D9  | 6  |
|        | A20 | D20 | 0  | B9  | D9  | 0  |
|        |     |     |    | D9  | B9  | 0  |
|        |     |     |    | D9  | D14 | 5  |
|        |     |     |    | B9  | B14 | 5  |
|        |     |     |    | D14 | B14 | 0  |
|        |     |     |    | B14 | D14 | 0  |
|        |     |     |    | A1  | A20 | 19 |
|        |     |     |    | D14 | D20 | 6  |
|        |     |     |    | A20 | D20 | 0  |
|        |     |     |    | D20 | A20 | 0  |